\documentclass[11pt]{article}

\usepackage{cite}
\usepackage{amsmath}
\usepackage{amsfonts}
\usepackage{amssymb}
\usepackage{epsf}

\def\ord#1{{\cal O}(#1)}
\def\bbox#1{\mbox{\boldmath$#1$}}

\begin{document}

\vspace*{0.3cm}

\begin{center}
\begin{tabular}{c}
\hline
\rule[-5mm]{0mm}{15mm}
{\Large \sf Double--Logarithmic Two--Loop Self--Energy}\\
\rule[-5mm]{0mm}{15mm}
{\Large \sf Corrections to the Lamb Shift}\\
\hline
\end{tabular}
\end{center}

\vspace{1cm}

\begin{center}
U.~D.~Jentschura$^{1),2)}$ and I. N\'andori$^{2),3)}$
\end{center}

\vspace{0.2cm}

\begin{center}
$^{1)}${\it Fakult\"{a}t Physik der Albert--Ludwigs--Universit\"{a}t,
Theoretische Quantendynamik,}\\
{\it Hermann--Herder--Stra\ss{}e 3, D--79104 Freiburg, Germany}\\[1ex]
$^{2)}${\it Institut f\"ur Theoretische Physik,}\\
{\it Technische Universit\"{a}t Dresden, 01062 Dresden, Germany} \\[1ex]
$^{3)}${\it Department of Theoretical Physics, University of Debrecen,
H-4032 Debrecen, Hungary}\\[1ex]
\end{center}

\vspace{1.3cm}

\begin{center}
\begin{minipage}{10.5cm}
{\underline{Abstract}}
Self-energy corrections involving logarithms of the parameter $Z\alpha$
can often be derived within a simplified approach, avoiding calculational
difficulties typical of the problematic non-logarithmic corrections
(as customary in bound-state quantum electrodynamics,
we denote by $Z$ the nuclear charge number, and by 
$\alpha$ the fine-structure
constant). For some logarithmic corrections, it is sufficient to consider
internal properties of the electron characterized by form factors.
We provide a detailed derivation of related self-energy ``potentials''
that give rise to the logarithmic corrections; these potentials are local
in coordinate space. We focus on the double-logarithmic two-loop
coefficient $B_{62}$ for P states and states with
higher angular momenta in hydrogenlike systems. We complement the discussion
by a systematic derivation of $B_{62}$ based on nonrelativistic 
quantum electrodynamics (NRQED). In particular, 
we find that an additional double
logarithm generated by the loop-after-loop diagram cancels when 
the entire gauge-invariant set of two-loop self-energy diagrams is
considered. This double logarithm is not contained in the 
effective-potential approach.
\end{minipage}
\end{center}

\vspace{1.3cm}

calculations and mathematical techniques in atomic and molecular physics;\\
Relativistic and quantum electrodynamic effects
in atoms and molecules.

\newpage

%
%
\section{Introduction} 

Lamb-shift measurements and related theoretical
calculations for bound atomic systems with increasing 
accuracy have historically provided accurate tests of quantum 
electrodynamics (QED), and the measurements have recently 
been improved in accuracy beyond previous 
limits~\cite{NiEtAl2000,GeLoHe2001,vWHoDr2001}. 
In order to account for a theoretical description, 
corrections of various physical origin (one-loop self-energy
and vacuum polarization, two-loop, 
and higher order radiative, recoil, radiative-recoil, nuclear-size 
corrections) have to be evaluated~\cite{EiGrSh2001}.

Here, we focus on logarithmic self-energy corrections which
are evaluated within the $Z\alpha$-expansion~\cite{SaYe1990}.
Within the analytic treatment,
self-energy radiative corrections can be taken into account by means of
a nonanalytic expansions in powers of the fine-structure constant
$\alpha$, the product of $Z\alpha$ and the logarithm 
$\ln[(Z\alpha)^{-2}]$ ($Z$ is the nuclear charge number). The 
expansion in powers of $\alpha$ corresponds to the loop-expansion 
in the framework of the usual perturbative treatment for QED. The 
higher order terms in powers of $Z\alpha$ and $\ln[(Z\alpha)^{-2}]$ 
are related to atomic-physics effects; they are referred to as the 
"binding corrections". 

The purpose of this investigation is twofold: first, to illustrate how
Lamb-shift ``potentials'' that give rise to the logarithmic corrections
can be derived within the context of bound-state QED, and 
second, to provide
a rigorous and detailed derivation of the $B_{62}$ double-logarithmic
two-loop self-energy coefficient for P
states and states with higher angular momenta.
The P-state coefficient $B_{62}$ has already 
appeared in the literature \cite{Ka1996}; 
however the derivation has been rather sketchy. 

%
%
\section{Modified Dirac Hamiltonian, One--Loop Corrections and $A_{41}$}
\label{section2}

It has been observed by many authors (e.g.~\cite{ItZu1980,Pa1999,JePa2002}) 
that
a rather important class of self-energy radiative effects for bound
states can be described by a modified Dirac Hamiltonian
($\hbar$ = $c$ = $\epsilon_0$ = 1),
\begin{eqnarray}
\label{dirac-ham}
H_D^{\rm (m)} &=& \bbox{\alpha} \, 
[\,\bbox{p} - e \, F_{1}(\Delta) \,\bbox{A}]
+ \beta \, m + e \, F_{1}(\Delta) \, \phi \nonumber\\[2ex]
&& \qquad + 
F_{2}(\Delta) \, \frac{e}{2 m} \, 
(\mathrm{i} \, \bbox{\gamma} \cdot \bbox{E} -
\beta \, \bbox{\sigma} \cdot \bbox{B} )\,,
\end{eqnarray}
which approximately describes an electron subject to an external
scalar potential $\phi \equiv \phi(r)$ and
an external vector potential $\bbox{A} \equiv \bbox{A(\bbox{r})}$ (the 
vector potential vanishes for a point nucleus that gives rise to
a static Coulomb potential; we may neglect the nuclear magnetic
field and the hyperfine structure). We have
\begin{equation}
e \, \phi(r) = e \, A_0(r) = - \frac{Z\alpha}{r}
\end{equation}
in coordinate space,
which corresponds to $\phi(\bbox{q}^2) = -4\pi Z\alpha/\bbox{q}^2$ 
in momentum space. In this article, following 
the commonly accepted convention, the function $\phi(r)$ and its Fourier
transform $\phi(\bbox{q}^2)$ are denoted by the same symbol
$\phi$. We avoid possible ambiguities by denoting with $\bbox{r}$
and $r$ the arguments in coordinate space and with 
$\bbox{q}$ or $\bbox{p}$ those in momentum space.
The argument $\Delta \equiv ({\partial}/{\partial \bbox{r}})^2$ 
of the electron form
factor $F_1$ in Eq.~(\ref{dirac-ham}) is to be interpreted as a Laplacian
operator acting on all quantities to the right, but not on the wave
function of the bound electron $\psi(\bbox{r})$.

Equation (\ref{dirac-ham}) entails a replacement of the 
binding Coulomb potential as
\[
e \, \phi(\bbox{r}) \to e \, F_{1}(\Delta) \, \phi(\bbox{r})
\]
and leads to a correction to the Coulomb potential
$\Delta V_{\rm C}(r)$ according to
\begin{equation}
\label{deltaVCcoor}
\Delta V_{\rm C}(r) = [F_{1}(\Delta) - 1] \, 
\left(-\frac{Z\alpha}{r}\right)
\end{equation}
in coordinate space, and 
\begin{equation}
\label{deltaVCmomen}
\Delta V_{\rm C}(\bbox{q}^2) = [F_{1}(-\bbox{q}^2)) - 1] \,
\left(-\frac{4\pi\,Z\alpha}{\bbox{q}^2}\right)
\end{equation}
in momentum space.
In first-order perturbation theory, this gives rise to the
following perturbative correction which we write down in coordinate and
momentum space,
\begin{eqnarray}
\label{E1}
\Delta E_1 \,
&=& \langle \psi \, | \Delta V_{\rm C}(r) | \, \psi \rangle
= \langle \psi \, | \, [F_{1}(\Delta) -1 ] \,
e \, \phi \, | \, \psi \rangle  \nonumber\\[2ex]
&=& \, \int {\rm d}^3 r \, \psi^{+}(\bbox{r}) \,
\left[ [F_1(\Delta) - 1] \, \left(\frac{-Z\alpha}{r}\right) \right]
\, \psi(\bbox{r}) \nonumber\\[2ex]
&=& \, \int \frac{{\rm d}^3 p}{(2\pi)^3} \, 
\int \frac{{\rm d}^3 p'}{(2\pi)^3} \, \psi^{+}(\bbox{p}') \,
\left[ [F_1(-\bbox{q}^2) - 1] \, 
\left(\frac{-4\pi Z\alpha}{\bbox{q}^2}\right) \right]
\, \psi(\bbox{p})\,,
\end{eqnarray}
with $\bbox{q} = \bbox{p}'-\bbox{p}$. An expansion of the electron form
factor $F_1$ in terms of its argument gives rise to higher-order
terms in the $Z\alpha$-expansion, because the atomic momentum is of
the order of $Z\alpha$ in natural units. Therefore, within the 
$Z\alpha$-expansion, it is admissible
to expand both the bound-state Dirac wavefunctions $\psi$ in powers
of $Z\alpha$ (the leading-order term is then the Schr\"{o}dinger
wavefunction), as well as the electron form factor in Eq.~(\ref{E1})
in powers of its argument.

The one-loop (1L) self-energy correction for S states
within the $Z\alpha$-expansion reads
\begin{eqnarray}
\label{coeff1L}
\Delta E^{({\rm 1L})}_{\rm SE} = 
\left(\frac{\alpha}{\pi}\right) \, (Z\alpha)^4 \, \frac{m}{n^3} \,
\left(A_{41} \, \ln[(Z\alpha)^{-2}] + A_{40} + \cal{R} \right)\,,
\end{eqnarray}
where the remainder $\cal{R}$ is of the order of $\ord{Z\alpha}$, $m$
is the electron mass and $n$ is the principal quantum number.

As indicated in Eq.~(\ref{E1}),
the form factor $F_1(\Delta)$ in momentum space assumes arguments
according to the replacement $\Delta \rightarrow - \bbox{q}^2 
\equiv -(\bbox{p}' - \bbox{p})^2$ in momentum space. 
With the convention $q^2 = q^{\mu} q_{\mu} =
(q^0)^2 - \bbox{q}^2$, the evaluation of the radiative 
corrections to the binding Coulomb field is mediated by space-like
virtual photons ($q^0=0$), and the momentum transfer can be written as:
$q^2 = -\bbox{q}^2 \equiv t$ (this is consistent with 
the conventions employed in~\cite{Re1972a,Re1972b}).

The form factor
$F_1(t)$ can be expanded in powers of $\alpha$, which corresponds
to the loop expansion. According to Eqs.~(1.2) and (1.20) 
of~\cite{Re1972a}, we have up to two-loop order: 
\begin{eqnarray}
F_1(t) = 1 +  \left(\frac{\alpha}{\pi}\right) \, F_1^{(2)}(t) 
+  \left(\frac{\alpha}{\pi}\right)^2  \, F_1^{(4)}(t) 
+  \ord{\alpha^3}
\end{eqnarray}
with
\begin{eqnarray}
\label{F11L}
F_1^{(2)}(t) &=& B(t) \, \ln\frac{\lambda}{m} 
+ {\cal F}_1^{(2)}(t) \,, \\[2ex]
\label{F12L}
F_1^{(4)}(t) &=& \frac{1}{2} \, B^2(t) \ln^2 \frac{\lambda}{m} 
+ \ln \frac{\lambda}{m} \, B(t) \, {\cal F}_1^{(2)}(t)
+ {\cal F}_1^{(4)}(t)\,,
\end{eqnarray}
where the ${\cal F}$ are infrared finite (i.e.~finite
in the limit $\lambda \to 0$), and the 
definition of the function $B(t)$ [see Eq. (1.18a) of 
\cite{Re1972a}] reads as follows,
\begin{eqnarray}
\label{Bt}
B(t) =  - \left[1 + \frac{t - 2\, m^2}{t (1-4 m^2/t)^{1/2}}
\ln\frac{(1-4 m^2/t)^{1/2} - 1}{(1-4 m^2/t)^{1/2} + 1}  \right]
= -\frac{t}{3 m^2} + \ord{t^2}\,.
\end{eqnarray}
In Eq.~(\ref{F11L}), $\lambda$ denotes the fictitious photon mass.
How should the problem of the infrared divergence of the 
form factors be interpreted in the context of bound-state QED?
The free electron can emit an infinite number
of infrared photons, because it may undergo transitions
between free states with infinitesimal energy differences.
However, this is not the case for a bound electron which 
has a discrete bound-state spectrum; energy levels
are separated from each
other by intervals of the order of $(Z\alpha)^2\,m$ 
(the energy level differences are determined by Schr\"{o}dinger theory). 
This leads to an infrared cutoff in bound-state QED of the order of 
$\lambda \approx (Z\alpha)^2\,m$. Therefore, we may replace
$\lambda \to  (Z\alpha)^2\,m$ for the determination of leading
logarithms of the Lamb shift. At some risk to over-simplification,
one may therefore argue that the infrared catastrophe is
avoided in a natural way for bound states. For the description
of bound states, we have $\ln(\lambda/m) \approx
- \ln[(Z\alpha)^{-2}]$ within logarithmic accuracy, i.e.~neglecting
non-logarithmic contributions which are given e.g.~by $A_{40}$ 
coefficients [see Eq.~(\ref{coeff1L})]. 

The focus of the current article is on double-logarithmic
corrections which are present from the first term on the right-hand side
of Eq.~(\ref{F12L}). Note that single-logarithmic two-loop corrections 
are not being considered in this article. Corrections of this latter type 
are generated, for example, by the {\em second} term on the right-hand side
of Eq.~(\ref{F12L}).   

At this point, it may be helpful to point out
that the cutoff of the infrared divergence of QED at the ``bound-state photon
mass'' $\lambda \to  (Z\alpha)^2\,m$ is consistent with 
the matching procedure that involves an explicit infrared
cutoff $\epsilon$ which can be interpreted
as an infrared cutoff for the bremsstrahlung 
spectrum~\cite{Pa1993,JePa2002,ItZu1980}. The procedure is described
in some detail in Eqs.~(32) -- (34) of~\cite{JePa2002}.
This matching procedure offers an alternative interpretation for the
infrared catastrophe: the infrared divergence crucially relies
on transitions between asymptotically free electron states. 
Any infinitesimally small additional interaction of the electrons within
that interferes with the emission
of bremsstrahlung will avoid the infrared catastrophe
and provide an infrared cutoff whose order-of-magnitude is determined
by the energy scale of the additional external field.

In combining the result (\ref{F11L}) with the 
expansion of $B(t)$ in powers of $t$, we reproduce the well-known
expression
\begin{equation}
F_1^{(2)}(t) = - \frac{t}{3 m^2} \, 
\left[ \ln \frac{\lambda}{m} + \frac{1}{8} \right] +
\ord{t^2}\,.
\end{equation}
Together with the definition of the modified Coulomb
potential in Eq.~(\ref{deltaVCmomen}) and the bound-state
``infrared-cutoff prescription'' $\lambda \to  (Z\alpha)^2\,m$, 
this leads to the following
one-loop (1L) self-energy potential
\begin{equation}
\label{pot1Lmomen}
\Delta V_{\rm C}^{\rm (1L)}(\bbox{q}^2) =
\frac{\alpha}{\pi} \, 
\left[ - \frac{-\bbox{q}^2}{3 m^2} \, (- \ln[(Z\alpha)^{-2}]) \right] \,
\left(- \frac{4 \pi Z \alpha}{\bbox{q}^2} \right)
= \frac{4 \alpha}{3 m^2} \, (Z\alpha) \, \ln[(Z\alpha)^{-2}] \,,
\end{equation}
in momentum space; this translates into a potential
\begin{equation}
\label{pot1Lcoor}
\Delta V_{\rm C}^{\rm (1L)}(r) =
\frac{4 \alpha}{3 \pi} \, (Z\alpha) \, \ln[(Z\alpha)^{-2}] \,
\frac{\delta^{(3)}(\bbox{r})}{m^2}
\end{equation}
in coordinate space. This potential can also be
found as Eq.~(2) of~\cite{IvKa1996},
given there without derivation. The first-order one-loop perturbation,
evaluated according to Eq.~(\ref{E1}), reads
\begin{equation}
\label{corr1L}
\Delta E^{\rm (1L)}_1 =
\langle \psi \, | \Delta V_{\rm C}^{\rm (1L)}(r) | \, \psi \rangle =
\frac{4 \alpha}{3 \pi} \, (Z\alpha)^4 \, \frac{m}{n^3} \,
\ln[(Z\alpha)^{-2}] \, \delta_{l0}\,.
\end{equation}
This correction
is nonvanishing only for S states ($l=0$), and it reproduces
the leading logarithmic $A_{41}$ coefficient as given in
Eq.~(\ref{coeff1L}). It may be interesting to point out that
since $|\psi(r=0)|^2 = (Z\alpha)^3 \, (m_{\rm r}^3/\pi) \, 
\delta_{l0}$, where $m_{\rm r}$ is the reduced mass of the system,
the correction (\ref{corr1L}) also has the correct reduced-mass
dependence (this is of relevance for systems like
positronium and pionium). In the limit of a large nuclear mass, we have
of course $m = m_{\rm r}$.

Note that the potential (\ref{pot1Lcoor}) is local in 
coordinate space. In contrast, the nonrelativistic (NR)
one-loop self-energy operator (as well as its relativistic counterpart
which assumes a slightly more complicated form) 
may be expressed in the length-gauge form as
[cf. Eq.~(29) of~\cite{JeSoIn2001})], 
\begin{equation}
\label{Sigma1L}
\Sigma^{\rm (1L)}_{\rm NR}(\bbox{r},\bbox{r}') = 
- \frac{2 \alpha}{3 \pi} \,
\int_0^\epsilon {\rm d}\omega \, \omega^3 \,
\bbox{r}' \, 
\left< \bbox{r}' \left| \frac{1}{H - E + \omega} \right| \bbox{r} \right>
\bbox{r}\,,
\end{equation}
where $\epsilon$ is the upper cutoff for the photon energy 
originally introduced in~\cite{Pa1993}.
The self-energy operator
(\ref{Sigma1L})
involves {\em two} spatial coordinates. The locality of the potential
(\ref{pot1Lcoor}) expresses the fact that the high-energy virtual
photons which mediate the form-factor corrections in Eq.~(\ref{dirac-ham})
act on a relativistic length scale given by the Compton wavelength 
of the electron which is smaller by one order of $Z\alpha$ 
than the atomic length scale given by the Bohr radius.

%
%
\section{Effective Local Potential for \\ Two--Loop Corrections and $B_{62}$}
\label{section3}

In combining the result (\ref{F12L}) with the
expansion of $B(t)$ in powers of $t$ [see Eq.~(\ref{Bt})] 
and the modified Coulomb
potential in Eq.~(\ref{deltaVCmomen}), and using
the bound-state
``infrared-cutoff prescription'' $\lambda \to  (Z\alpha)^2\,m$,
we obtain the following
two-loop (2L) self-energy potential
\begin{eqnarray}
\label{pot2Lmomen}
\Delta V_{\rm C}^{\rm (2L)}(\bbox{q}^2) &=&
\left( \frac{\alpha}{\pi} \right)^2 \,
\frac{1}{2} \, \left( \frac{\bbox{q}^2}{3\,m^2} \right)^2 \,
\, \ln^2[(Z\alpha)^{-2}] \, 
\left(- \frac{4 \pi \, Z \alpha}{\bbox{q}^2}\right)
\nonumber\\[2ex]
&=& \left(\frac{\alpha}{\pi}\right)^2 \,
\frac{1}{18} \, \ln^2[(Z\alpha)^{-2}] \, \frac{4 \pi \, Z\alpha}{m^4}
\left(-\bbox{q}^2 \right) \,.
\end{eqnarray}
This correction has previously appeared
as Eq.~(3) of~\cite{Ka1996}, without a detailed derivation.
After Fourier transformation, we have
\begin{equation}
\label{pot2Lcoor}
\Delta V_{\rm C}^{\rm (2L)}(r) =
\frac{2}{9} \, \left(\frac{\alpha}{\pi}\right)^2 \,
\ln^2[(Z\alpha)^{-2}] \, \frac{\pi \, 
\Delta \delta^{(3)}(\bbox{r})}{m^4}\,,
\end{equation}
which is a highly singular potential in coordinate space.
Its expectation value on S states diverges, giving rise 
to a further logarithm, and we will
not discuss here the associated problems, which have
recently attracted remarkable 
attention~\cite{MaSa1998b,GoLaNePlSo1999,MaSt2000,Ye2000,Ye2001,
MaSoSt2001,MaSt2001,Pa2001}.

\begin{figure}[htb]
\begin{center}
\begin{minipage}{12cm}
\centerline{\mbox{\epsfysize=7.7cm\epsffile{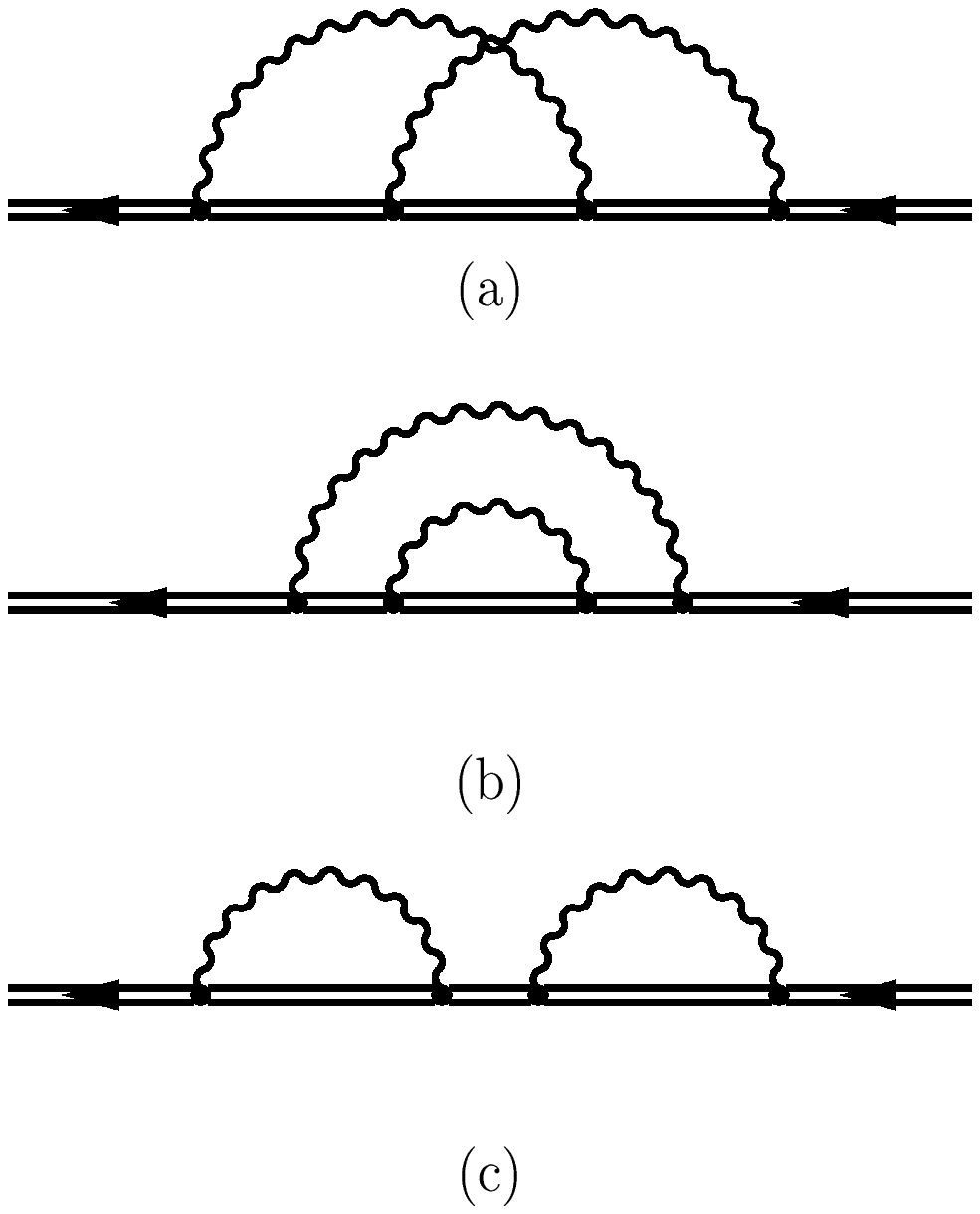}}}
\caption{\label{fig1} The crossed (a), rainbow (b) and the
loop-after-loop diagram (c) which contribute to the two-loop
self-energy for a bound electron. The propagator of the bound
electron is denoted by a double line.}
\end{minipage}
\end{center}
\end{figure}

The first-order perturbation, evaluated according to
Eq.~(\ref{E1}), reads
\begin{eqnarray}
\label{Laplace}
\Delta E^{\rm (2L)}_1 = \left(\frac{\alpha}{\pi}\right)^2
\frac{2}{9}  \frac{\pi Z\alpha}{m^4} \,  
\ln^2\left[(Z\alpha)^{-2}\right] \,
\left. \Delta \left[\left| \phi_{n,l=1,m}(\bbox{r}) \right|^2 \right]
\right|_{r=0}.
\end{eqnarray} 
In Eq.~(\ref{Laplace}), the Laplacian operator acts on a Schr\"odinger 
P wavefunction. The following analytic result
\begin{eqnarray}
\label{Laplace2}
\left. \Delta \left[\left| \phi_{n,l=1,m}(\bbox{r}) \right|^2 \right]
\right|_{r=0} = \frac{2}{3\pi}
\left[ (Z \alpha)^5 m^5 \right] \frac{n^2-1}{n^5}\,,
\end{eqnarray} 
where $n$ is the principal quantum number, has previously appeared in 
the literature (e.g.~\cite{Ka1996,Pa1999}).
Within the current investigation, we would like to present 
a complete derivation of the analytic expression
for this matrix element in the Appendix~\ref{appa}.
Finally, we rewrite the energy correction in the form
\begin{eqnarray}
\label{E2L1}
\Delta E^{\rm (2L)}_1 = \left(\frac{\alpha}{\pi}\right)^2
\frac{(Z\alpha)^6 m}{n^3} \, 
\ln^2\left[(Z\alpha)^{-2}\right] \,
\frac{4}{27} \, \frac{n^2-1}{n^2}\,.
\end{eqnarray} 
This double-logarithmic correction originates solely
from the two-loop $F_1$ form factor of the electron.
This corresponds to the diagrams in Fig.~\ref{fig1} (a) and (b).
To complete the gauge-invariant set, the loop-after-loop
diagram in Fig.~\ref{fig1} (c) should also be taken into
consideration. 

The diagram in Fig.~\ref{fig1} (c) gives rise to a 
``second-order perturbation'' involving to one-loop self
energies as first-order perturbations (the ``irreducible part''
of the diagram), supplemented by a further term
involving the derivative of the bound electron's Green function
(the ``reducible part''). 
The correction is known to read (see e.g.~\cite{MoPlSo1998})
\begin{equation}
\label{diagc}
\left< \overline\psi \left| \Sigma^{\rm (1L)}_{\rm R}(E) \,
\left( \sum_{\psi' \neq \psi} \frac{|\psi'\rangle\,
\langle \overline\psi' |}{E - E_{\psi'}} \right) \,
\Sigma^{\rm (1L)}_{\rm R}(E) \right| \psi \right>
+
\left< \overline\psi \left| 
\Sigma^{\rm (1L)}_{\rm R} (E) \right| \psi \right>
\left< \overline\psi \left| \frac{\mathrm d}{{\mathrm d} E}
\Sigma^{\rm (1L)}(E) \right| \psi \right>\,,
\end{equation}
where $\Sigma^{\rm (1L)}_{\rm R}(E)$ is the renormalized
relativistic one-loop self-energy operator, and $E$
is the energy of the electron in the state $|\psi\rangle$. 
{\em Within the effective-potential approach},
the one-loop potential (\ref{pot1Lcoor})
describes the two one-loop self-energy insertions
in the first term of (\ref{diagc}).
The potential (\ref{pot1Lcoor})
involves a Dirac delta-function in coordinate space
that vanishes on P states,
and consequently it can be argued that
no further double-logarithmic corrections 
originate from this term ({\em but} see the discussion in
Secs.~\ref{section4} and~\ref{section5}).

The second term in (\ref{diagc}),
which involves the derivative of the self-energy operator
with respect to its argument [see also Eq.~(2.6) of~\cite{Pa1993} 
or Eq.~(2) of~\cite{Ka1996}] and constitutes the reducible
part of the diagram in Fig.~\ref{fig1} (c),
does not give rise to any further double logarithm, either.
The first factor $\left< \overline\psi \left|
\Sigma^{\rm (1L)}_{\rm R} (E) \right| \psi \right>$
does not create any logarithm for P states
in the order of $\alpha \, (Z\alpha)^4$. The second factor,
which contains the derivative of the self-energy operator,
is not separately gauge invariant, and consequently, there exists 
no ``effective potential'' which could be
inserted for this term. This is in itself a rather
unsatisfactory situation for the effective-potential approach. 
However, it is possible to analyze 
the logarithm which is generated by the nonrelativistic
photon integration region in this term. Consider the 
nonrelativistic ``velocity-gauge'' form of (\ref{Sigma1L}) and differentiate
with respect to the energy,
\begin{equation}
\left< \overline\psi \left| \frac{\mathrm d}{{\mathrm d} E}
\Sigma^{\rm (1L)}(E) \right| \psi \right>_{\rm NR}
= - \frac{2 \alpha}{3 \pi} \,
\int_0^\epsilon {\rm d}\omega \, \omega \,
\left< \phi \left| \frac{\bbox{p}}{m} 
\left( \frac{1}{H - E + \omega} \right)^2 
\frac{\bbox{p}}{m} \right| \phi \right>\,,
\end{equation}
where $\phi$ is the nonrelativistic (Schr\"{o}dinger)
wavefunction,
There is only a {\em single} logarithm 
$\ln[\epsilon/(Z\alpha)^2 m]$ generated in the 
integration region $\omega \in [(Z\alpha)^2\,m, \epsilon]$
which may be extracted by replacing $1/(H - E + \omega) \to
1/\omega$. The logarithmic term is proportional to the matrix element
$\langle \phi | (\bbox{p}^2/m^2) | \phi \rangle$, which is 
finite on P states. Consequently, no further double logarithms
arise from the second term of (\ref{diagc}).

%
%

The two-loop effect for P states is usually characterized by the 
following semi-analytic expansion in powers of $Z\alpha$
[cf.~Eq.~(\ref{coeff1L})],
\begin{eqnarray}
\label{coeff2L}
\Delta E^{\rm (2L)}_{\rm SE} = 
\left(\frac{\alpha}{\pi}\right)^2 \, (Z\alpha)^4 \,
\frac{m}{n^3} \,\left( B_{40} + (Z \alpha)^2 \, 
[B_{62} \ln^{2}(Z\alpha)^{-2}\right. \nonumber\\[2ex]
\hskip 5cm \left. + B_{61} \ln(Z\alpha)^{-2} 
+ B_{60}] + \cal{R} \right),
\end{eqnarray}
where the remainder $\cal{R}$ is order $\ord{Z\alpha}^{3}$.
Using Eq.~(\ref{E2L1}),
one can immediately read off the two-loop double logarithmic 
spin-independent coefficient
\begin{eqnarray}
\label{result}
B_{62}(n,l=1) = \frac{4}{27} \, \frac{n^2-1}{n^2}\,.
\end{eqnarray} 
We confirm the result obtained for this 
correction in Ref.~\cite{Ka1996}. 

%
%
\section{Double Logarithms and the Loop--After--Loop
  Diagram}
\label{section4}

In the previous section, we have seen that {\em within the 
effective-potential approach}, no double logarithm
originates in the order $(Z\alpha)^6$ from the loop-after-loop
diagram in Fig.~\ref{fig1} (c). This is because, within this
approach,  we insert the delta-like
local potential (\ref{pot1Lcoor}) for the two one-loop self-energies
in the first term of (\ref{diagc}). 

However, if we consider 
the diagram in Fig.~\ref{fig1} (c) within the Coulomb gauge and
formulate the contribution due to low-energy virtual photons, then
we obtain for the irreducible part the expression
\begin{equation}
\label{addouble}
\Delta E_{\rm LAL} = - \left< \phi_{n,1,m} \left| \Sigma^{\rm (1L)}_{\rm NR} \,
\, \left( \frac{1}{H - E} \right)' \, \Sigma^{\rm (1L)}_{\rm NR} \right|
\phi_{n,1,m} \right>\,,
\end{equation}
where the nonrelativistic self-energy operator is given by Eq.~(\ref{Sigma1L}),
and $\phi_{n,1,m}$ is the Schr\"{o}dinger P wavefunction [see also 
Eq.~(\ref{schroedi})], the prime denotes the reduced Green function,
and $E$ is the energy of the $n$P state (``LAL'' = loop-after-loop). 
The double-logarithmic term $\Delta E^{2{\rm log}}_{\rm LAL}$
originating from (\ref{addouble}) reads
\begin{eqnarray}
\Delta E^{2{\rm log}}_{\rm LAL} &=& -
\frac{4}{9} \, \left( \frac{\alpha}{\pi} \right)^2 \,
\ln^2\left[\frac{\epsilon}{(Z\alpha)^2 m}\right]\,
\nonumber\\[2ex]
&& \times \left< \phi_{n,1,m} \left| \, \frac{\bbox{p}}{m} \, (H - E) \, 
\frac{\bbox{p}}{m} \,
\, \left( \frac{1}{H - E} \right)' \, 
\frac{\bbox{p}}{m} \, (H - E) \, \frac{\bbox{p}}{m} \, \right| 
\phi_{n,1,m} \right>\,.
\end{eqnarray}
In order to obtain this result, the denominator
of the Green function $H - E + \omega$ has been expanded
in powers of $H - E$ within the integration region
$\omega \in [(Z\alpha)^2\,m, \epsilon]$.
Using the commutator relation 
\begin{equation}
\label{aba}
ABA = \frac12 \, ([A,[B,A]] + A^2 B + B A^2)\,,
\end{equation}
with $A = \bbox{p}/m$ and $B = H - E$,
the matrix element can be rewritten in a much simpler fashion, and the 
double-logarithmic term becomes
\begin{equation}
\label{Trewritten}
\Delta E^{2{\rm log}}_{\rm LAL} = -
\frac{1}{9} \, \left( \frac{\alpha}{\pi} \right)^2 \,
\ln^2\left[\frac{\epsilon}{(Z\alpha)^2 m}\right]\, \frac{1}{m^4} \,
\left< \phi_{n,1,m} \left| \, \bbox{p}^2 \, (H - E) \, \bbox{p}^2 
\right| \phi_{n,1,m} \right>\,.
\end{equation}
We have
\begin{equation}
\label{matrixelem}
\left< \phi_{n,1,m} \left| \, \bbox{p}^2 \, (H - E) \, \bbox{p}^2 
\right| \phi_{n,1,m} \right> = \frac{(Z\alpha)^6 m^5}{n^3} \,
\left( \frac{4}{5} - \frac{8}{15 n^2} \right) \,,
\end{equation}
Note that for S states, the above matrix element is divergent, and
a regularization of the matrix element gives rise to an additional
(triple) logarithm $B_{63}$.
With the natural ultraviolet cutoff $\epsilon \approx m$ for 
nonrelativistic QED, we obtain from (\ref{Trewritten}) 
and (\ref{matrixelem}) the following double-logarithmic contribution,
\begin{equation}
\label{ELL2log}
\Delta E^{2{\rm log}}_{\rm LAL}(n, l=1)  = -
\left( \frac{\alpha}{\pi} \right)^2 \, \frac{(Z\alpha)^6 \, m}{n^3} \,
\ln^2[(Z\alpha)^{-2}] \, \,
\left( \frac{4}{45} - \frac{8}{135 n^2} \right) \,.
\end{equation}
Note that the presence of an additional double-logarithmic term originating
from the loop-after-loop diagram in Fig.~\ref{fig1} (c) in the Coulomb
gauge does not imply that
the result given in (\ref{result}) for the total value of 
$B_{62}$ is necessarily
incomplete, but it means that additional double logarithms have to 
expected if, e.g., this diagram is treated numerically, and numerical
and analytic results are compared. 
For S states, an additional contribution to the triple logarithm
$B_{63}$ originating from the loop-after-loop diagram was found 
in~\cite{MaSa1998b,Ye2000,Ye2001}, but the result originally 
obtained in~\cite{Ka1996}
for the {\em total} value of $B_{63}$ was confirmed in~\cite{MaSt2000,Pa2001}.
In the following section, we will 
derive the result (\ref{result}) by an independent calculation
which includes the entire gauge-invariant 
set of the diagrams in Fig.~\ref{fig1} in a rigorous way.

%
%
\section{Derivation Based on NRQED}
\label{section5}

We start from the expression [see Eq.~(16) of~\cite{Pa2001}],
\begin{eqnarray}
\label{NRQED}
\Delta E_{\rm NRQED} &=& - \left( \frac{2 \, \alpha}{3 \,\pi\,m^2} \right) \,
\int_0^{\epsilon_1} {\rm d}\omega_1 \, \omega_1 \,
\int_0^{\epsilon_1} {\rm d}\omega_2 \, \omega_2 \, \nonumber\\[1ex]
& & \left\{ 
\left< p^i \, \frac{1}{H - E + \omega_1} \, p^j \,
\frac{1}{H - E + \omega_1 + \omega_2} \, p^i \, 
\frac{1}{H - E + \omega_2} \, p^j \right> \right. \nonumber\\[1ex]
& & + \frac{1}{2} \,
\left< p^i \, \frac{1}{H - E + \omega_1} \, p^j \,
\frac{1}{H - E + \omega_1 + \omega_2} \, p^j \, 
\frac{1}{H - E + \omega_2} \, p^i \right> \nonumber\\[1ex]
& & + \frac{1}{2} \,
\left< p^i \, \frac{1}{H - E + \omega_2} \, p^j \,
\frac{1}{H - E + \omega_1 + \omega_2} \, p^j \, 
\frac{1}{H - E + \omega_1} \, p^i \right>
\nonumber\\[1ex]  
& & + 
\left< p^i \, \frac{1}{H - E + \omega_1} \, p^i \,
\left( \frac{1}{H - E} \right)' \, p^j \, 
\frac{1}{H - E + \omega_2} \, p^i \right> 
\nonumber\\[1ex]  
& & - \frac{1}{2} \,
\left< p^i \, \frac{1}{H - E + \omega_1} \, p^i \right> \,
\left< p^j \, \left( \frac{1}{H - E + \omega_2} \right)^2 \, p^i \right>
\nonumber\\[1ex]
& & - \frac{1}{2} \,
\left< p^i \, \frac{1}{H - E + \omega_2} \, p^i \right> \,
\left< p^j \, \left( \frac{1}{H - E + \omega_1} \right)^2 \, p^i \right>
\nonumber\\[1ex]
& & - m \,
\left< p^i \, \frac{1}{H - E + \omega_1} \, 
\frac{1}{H - E + \omega_2} \, p^i \right>
\nonumber\\[1ex]
& & \left. - \frac{m}{\omega_1 + \omega_2} \,
\left< p^i \, \frac{1}{H - E + \omega_2} \, p^i \right>
- \frac{m}{\omega_1 + \omega_2} \,
\left< p^i \, \frac{1}{H - E + \omega_1} \, p^i \right>
\right\}\,.
\end{eqnarray}
All of the matrix elements are evaluated on the reference state
$|\phi\rangle$, which can be taken as the Schr\"{o}dinger wave
function.

Within the $\epsilon$-method~\cite{Pa1993,JePa1996,JePa2002}, 
we extract those divergent
contributions from (\ref{NRQED})  
that involve double logarithms
$\alpha^2 \, (Z\alpha)^6 \, \ln^2[\epsilon/(Z\alpha)^2 m]$ (we may put 
$\epsilon = \epsilon_1 = \epsilon_2$ for simplicity). 
These logarithms correspond to the
ultraviolet divergence of NRQED and are generated
by the contributions of two infrared photons
($\omega_1 < \epsilon, \omega_2 < \epsilon$).
The divergences in $\epsilon$ necessarily cancel at the
end of the calculation due to contributions proportional
to $\ln (m/\epsilon) \, \ln[\epsilon/(Z\alpha)^2 m]$
which are generated by
intermediate integration regions ($\omega_1 > \epsilon,
\omega_2 < \epsilon$), and by terms proportional to
$\ln^2 (m/\epsilon)$ originating from high-energy virtual photons
($\omega_1 > \epsilon, \omega_2 > \epsilon$).
The latter terms correspond to
the infrared divergent terms proportional to $\ln^2 (\lambda/m)$
of the electron form factors.
For a discussion of the related cancellations in the
context of the $\epsilon$ method, we refer to~\cite{Pa1993} and
the Appendix of~\cite{JePa2002}. For the double logarithms,
the dependence on $\epsilon$ cancels between the low-energy,
the intermediate and the high-energy regions
according to $\ln^2[\epsilon/(Z\alpha)^2 m] +
2 \ln (m/\epsilon) \, \ln[\epsilon/(Z\alpha)^2 m] +
\ln^2 (m/\epsilon) = \ln[(Z\alpha)^{-2}]$.

There are nine terms in curly brackets on the right-hand side
of Eq.~(\ref{NRQED}) which we would like to denote
by ${\mathcal T}_1$ -- ${\mathcal T}_9$. These fall quite naturally 
into six groups, giving rise to six double logarithms ${\mathcal L}_1$
-- ${\mathcal L}_6$ according to the following correspondence:
\begin{itemize}
\item ${\mathcal T}_1 \to {\mathcal L}_1$,
\item ${\mathcal T}_2 + {\mathcal T}_3 \to {\mathcal L}_2$,
\item ${\mathcal T}_4 \to {\mathcal L}_3$,
\item ${\mathcal T}_5 + {\mathcal T}_6 \to {\mathcal L}_4$,
\item ${\mathcal T}_7 \to {\mathcal L}_5$,
\item ${\mathcal T}_8 + {\mathcal T}_9 \to {\mathcal L}_6$.
\end{itemize}
After an integration in the logarithmic region 
$\omega_1 \in [(Z\alpha)^2 m, \epsilon_1]$ and
$\omega_2 \in [(Z\alpha)^2 m, \epsilon_2]$, the logarithmic
contributions can be expressed by matrix elements,
evaluated on the reference state, according to the following
formulas (again,
we put for simplicity $\epsilon = \epsilon_1 = \epsilon_2$):
\begin{eqnarray}
\label{L1}
{\mathcal L}_1 &=& \left( \frac{\alpha}{\pi} \right)^2 \,
\ln^2\left[ \frac{\epsilon}{(Z\alpha)^2} \right] \,
\frac{4 \langle p^i \, (H-E) \, p^i \, \bbox{p}^2 \rangle}
  {9 \, m^4} \,, 
\\[2ex]
\label{L2}
{\mathcal L}_2 &=& \left( \frac{\alpha}{\pi} \right)^2 \,
\ln^2\left[ \frac{\epsilon}{(Z\alpha)^2} \right] \,
 \frac{2 \, \langle p^i \, p^j \, (H-E) \, p^j \, p^i \rangle
   - 4 \langle p^i \, (H-E) \, p^i \, \bbox{p}^2 \rangle}{9 \, m^4} \,,
\\[2ex]
\label{L3}
{\mathcal L}_3 &=& \left( \frac{\alpha}{\pi} \right)^2 \,
\ln^2\left[ \frac{\epsilon}{(Z\alpha)^2} \right] \,
 \frac{- \langle \bbox{p}^2 \, (H-E) \, \bbox{p}^2 \rangle}{9 \, m^4} \,,
\\[2ex]
{\mathcal L}_4 &\propto& \langle p^i \, (H-E) \, p^i \rangle = 0\,,
\\[2ex]
{\mathcal L}_5 &=& \left( \frac{\alpha}{\pi} \right)^2 \,
\ln^2\left[ \frac{\epsilon}{(Z\alpha)^2} \right] \,
 \frac{4 \langle p^i \, (H-E)^2 \, p^i \rangle}{9 \, m^3} \,,
\\[2ex]
\label{L6}
{\mathcal L}_6 &=& \left( \frac{\alpha}{\pi} \right)^2 \,
\ln^2\left[ \frac{\epsilon}{(Z\alpha)^2} \right] \,
 \frac{- 4 \langle p^i \, (H-E)^2 \, p^i \rangle}{9 \, m^3} \,.
\end{eqnarray}
All of these matrix elements are finite when evaluated
on P states and on states with higher angular momenta.
In deriving these results, use is made of the integrals 
$I_1$ -- $I_3$ listed in Appendix~\ref{appb}. In particular, $I_1$ is used
in deriving ${\mathcal L}_1$, $I_2$ is used in deriving ${\mathcal L}_2$,
and ${\mathcal L}_6$ can be derived using $I_3$. The double
logarithm ${\mathcal L}_3$ 
corresponds to Eq.~(\ref{Trewritten}). Summing all
contributions ${\mathcal L}_1$ -- ${\mathcal L}_6$, we obtain
\begin{eqnarray}
{\mathcal L} &=& \sum_{i=1}^6 {\mathcal L}_i 
\nonumber\\[2ex]
&=& \left( \frac{\alpha}{\pi} \right)^2 \,
\ln^2\left[ \frac{\epsilon}{(Z\alpha)^2} \right] \,
 \frac{2 \, \langle p^i \, p^j \, (H-E) \, p^j \, p^i \rangle
  - \langle \bbox{p}^2 \, (H-E) \, \bbox{p}^2 \rangle}{9 \, m^4}
\nonumber\\[2ex]
&=& \left( \frac{\alpha}{\pi} \right)^2 \,
\ln^2\left[ \frac{\epsilon}{(Z\alpha)^2} \right] \,
  \frac{2 \pi \, \langle \Delta \delta^{(3)}(\bbox{r}) \rangle}{9 \, m^4}
\end{eqnarray}
in agreement with formulas (\ref{pot2Lcoor}) and (\ref{Laplace}).
Here, use is made of the equality
\begin{equation}
\label{pipjpjpi}
\langle p^i \, p^j \, (H-E) \, p^j \, p^i \rangle =
\pi \, (Z\alpha) \, \langle \Delta \delta^{(3)}(\bbox{r}) \rangle
+ \frac{1}{2} \,  \langle \bbox{p}^2 \, (H-E) \, \bbox{p}^2 \rangle\,,
\end{equation}
which is valid for P states and states with higher 
angular momenta and can be derived using the commutator 
relation (\ref{aba}). We thereby confirm that the additional
double logarithm ${\mathcal L}_3$ generated by the loop-after-loop diagram
Fig.~\ref{fig1} (c) is cancelled by an additional contribution
from ${\mathcal L}_2$ according to Eqs.~(\ref{L2}) and (\ref{pipjpjpi}).

As a byproduct of the current investigation,
we obtain the rigorous result that $B_{62}$ vanishes for states with
higher angular momenta $l \geq 2$. This is because the
expectation value of the effective potential~(\ref{pot2Lcoor}),
when evaluated on hydrogenic D, F, G,$\dots$ states, vanishes:
states with higher angular momenta behave as $r^l$ for small
$r$, where $l$ is the angular momentum. We thereby confirm a
statement made in~\cite{Ka1996} [following Eq.~(5) {\em ibid.}]
where it was pointed out
that a formula analogous to (\ref{Laplace}) holds for
all states with $l\neq1$ [see the text following Eq.~(5) {\em ibid}.].

%
%
\section{Results and Conclusions}

The results of the current investigation can be summarized as follows:
In Sec.~\ref{section2}, 
we attempt to clarify the derivation and physical origin of effective
potentials~\cite{Ka1996} used for the approximate description 
of self-energy corrections
in leading logarithmic accuracy,
and to provide
a more detailed derivation of known double-logarithmic
corrections to the Lamb shift.
In Sec.~\ref{section3}, restricting the discussion to P states and states
with higher angular momenta, we rederive, within the
effective-potential approach, known results~\cite{Ka1996}
for the leading spin-independent double logarithm for P states as given by
the $B_{62}$ coefficient [see Eq.~(\ref{result})].
In Sec.~\ref{section4}, we show that nonvanishing double logarithms have
to be expected from the loop-after-loop diagram if this non-gauge invariant
term is treated separately (e.g., within a numerical
evaluation). By contrast, within the effective-potential approach,
the double logarithm for this diagram {\em vanishes}
(see the entry in column 2, row 4 of Table 1 of~\cite{Ka1996}).
In Sec.~\ref{section5}, we show that a rigorous derivation
of $B_{62}$ based on the entire gauge-invariant set of diagrams in 
Fig.~\ref{fig1} confirms the result (\ref{result}) for the
{\em total} value of $B_{62}$. In particular, the 
additional double logarithm originating from the loop-after-loop diagram 
cancels when the contributions of all diagrams are added,
and $B_{62}$ vanishes for all states with angular momenta $l > 1$. 

A reliable understanding of the problematic two-loop
corrections is important for the determination of fundamental
constants from precision spectroscopy~\cite{MoTa2000}.
We would also like to stress that analytic calculations,
even in the low--$Z$ region, could be supplemented by accurate
numerical evaluations in the near future. Recently,
a complete evaluation of the two-loop self-energy
effect for high--$Z$
has been reported~\cite{YeSh2001}. A comparison of the
numerical to the analytic results represents a crucial test 
for both methods~\cite{JeMoSo1999}. In order to provide for 
a reliable comparison of numerical vs.~analytic results, it is helpful 
to thoroughly analyze and understand the
logarithmic terms from each one of the diagrams in Fig.~\ref{fig1}.
As outlined in Sec.~5 of~\cite{JePa2002},
the most accurate theoretical predictions for the energy levels
can be obtained using
a combination of analytic and numerical results.

\section*{Acknowledgments}

The authors acknowlegde many insightful
discussions with K. Pachucki.
This work has been supported by the Deutscher Akademischer
Austauschdienst (DAAD). 

\appendix
 
%
%
\section{Appendix A: Analytic Evaluation of a Matrix Element}
\label{appa}

In this appendix,
we discuss the derivation of the expression (\ref{Laplace2}) 
\[
\left. \Delta \left[\left| \phi_{n,l=1,m}(\bbox{r}) \right|^2 \right]
\right|_{r=0} 
\]
for hydrogenic P states.
In Eq.~(\ref{Laplace}), the Laplacian operator
acts on nonrelativistic, Schr\"odinger wavefunctions,
which are given by 
\begin{equation}
\label{schroedi}
\phi_{n,l=1,m}(\bbox{r}) = R_{n 1}(r) Y_{1 m}(\theta, \phi)\,,
\end{equation} 
where $R_{n 1}(r)$ is the radial component, $Y_{1 m}(\theta, \phi)$ is 
the spherical harmonics with the polar-coordinates $r$, $\theta$ and $\phi$ 
and with quantum numbers ($n, l=1, m$). Since the quantum number $l=1$
than the magnetic quantum number can be $m=0$ and $m=1,-1$. For the sake of
simplicity we consider the $m=0$ case,
\begin{eqnarray}
Y_{1 m=0}(\theta, \phi) = \left(\frac{3}{4\pi} \right)^{1/2} \cos\theta.
\end{eqnarray} 
The Laplacian in (\ref{Laplace}) can be written in polar-coordinates as  
\begin{eqnarray}
\Delta \equiv \Delta_r + \Delta_{\theta,\phi} \equiv 
\left( \frac{1}{r^2} \, \frac{\partial}{\partial r} \,
r^2 \, \frac{\partial}{\partial r} \right)
+ \frac{1}{r^2} \left( \frac{1}{\sin\theta} 
\frac{\partial}{\partial \theta} \,
\left(\sin\theta \, \frac{\partial}{\partial \theta} \right) + 
\frac{1}{\sin^2\theta} \frac{\partial^2}{\partial \phi^2}\right)\,,
\end{eqnarray} 
where $\Delta_r$ corresponds to the radial component and 
$\Delta_{\theta,\phi}$ stands for the angular-dependent part of the
Laplacian operator. One easily obtains 
\begin{eqnarray}
\label{performed}
\Delta \left[\left| \phi_{n,l=1,m=0}(\bbox{r}) \right|^2 \right] =
\Delta_r R^2_{n 1} \, \frac{3}{4\pi} \, \cos^2\theta + 
R^2_{n 1} \, \frac{3}{4\pi} \,
\frac{2}{r^2} (1- 3 \cos^2 \theta). 
\end{eqnarray} 
The final result (\ref{Laplace2}) should be
independent of the angle $\theta$, i.e.~independent of the
spatial direction in which the origin is approached, and 
independent of the magnetic quantum number. 
Therefore, we may postulate that the $\theta$-dependent terms in
(\ref{performed}) mutually cancel. Alternatively, we observe that
since (\ref{Laplace2}) should be 
independent of the angle $\theta$, so that 
so that we may set $\theta = \pi/2$.
Reading off the $\theta$-independent part of (\ref{performed}), 
the following result can be obtained:
\begin{eqnarray}
\left. \Delta \left[\left| \phi_{n,1,0}(\bbox{r}) \right|^2 \right]
\right|_{r \to 0} =
\frac{3}{4\pi} 
\left.\left(\frac{2}{r^2} R^2_{n 1}\right)\right|_{r \to 0}. 
\end{eqnarray} 
The radial component of the Schr\"odinger wavefunction for hydrogenlike
P states ($R_{n 1}$) is defined by the associated Laguerre polynomials 
($L^{3}_{n+1}$) which read
\begin{eqnarray}
R_{n 1}(r) &=& - \left(\frac{(n-2)!}{(n+1)!^3 \, (2 n)!}\right)^{1/2}
\left(\frac{2}{n a_B}\right)^{5/2} \, r \, 
\exp\left(\frac{-r}{n a_B}\right) \, 
\, L^{3}_{n+1}\left(\frac{2 r}{n a_B}\right)\,, \nonumber\\[3ex]
L^{3}_{n+1}(\rho) &=& 
\frac{\partial^3}{\partial \rho^3} \sum^{n+1}_{j=0} (-1)^j 
\, {n+1 \choose j} \, \frac{(n+1)!}{j!} \, \rho^j\,,
\end{eqnarray} 
where the Bohr radius is $a_B = 1/(Z \alpha m)$. Using this relation,
it is straightforward to obtain
\begin{eqnarray}
\left. \Delta \left[\left| \phi_{n,1,0}(\bbox{r}) \right|^2 \right]
\right|_{r \to 0} = \frac{2}{3\pi} 
\left[ (Z \alpha)^5 m^5 \right] \frac{n^2 - 1}{n^5}\,,
\end{eqnarray} 
which is equivalent to Eq.~(\ref{Laplace2}).

%
%
\section{Appendix B: Double--Logarithmic Integrals}
\label{appb}

In this Appendix, we provide the results for certain integrals
which may be used in order to extract the double-logarithmic
contributions of order $(Z\alpha)^6 \, \ln^2 [\epsilon/(Z\alpha)^2]$
from the NRQED two-loop self-energy (\ref{NRQED}). We have two photon
energies $\omega_1$ and $\omega_2$ and denote arbitrary matrix elements
of the various occurrences of the 
operator $H-E$, scaled by $(Z\alpha)^2$,  by the symbols
$A_1$, $A_2$ and
$A_3$, respectively. The symbol $\sim$ in this Appendix is meant to 
indicate that only the double-logarithmic terms of order
$(Z\alpha)^6$ are selected. We have,
\begin{eqnarray}
\label{I1}
I_1 &=& \int_0^{\epsilon_1} {\rm d}\omega_1 \, \omega_1 \,
\int_0^{\epsilon_2} {\rm d}\omega_2 \, \omega_2 \,
\frac{(Z\alpha)^2}{\omega_1 + A_1 \, (Z\alpha)^2} \,
\frac{1}{\omega_1 + \omega_2 + A_2 \, (Z\alpha)^2} \,
\frac{(Z\alpha)^2}{\omega_2 + A_3 \, (Z\alpha)^2} 
\nonumber\\[2ex]
&\sim& - \frac{1}{2} \, (Z\alpha)^6 \,
\ln \frac{\epsilon_1}{(Z\alpha)^2} \,
\ln \frac{\epsilon_2}{(Z\alpha)^2} \,
\left(A_1 + A_3\right)\,,
\\[3ex]
\label{I2}
I_2 &=& \int_0^{\epsilon_1} {\rm d}\omega_1 \, \omega_1 \,
\int_0^{\epsilon_2} {\rm d}\omega_2 \, \omega_2 \,
\frac{(Z\alpha)^2}{\omega_1 + A_1 \, (Z\alpha)^2} \,
\frac{1}{\omega_1 + \omega_2 + A_2 \, (Z\alpha)^2} \,
\frac{(Z\alpha)^2}{\omega_2 + A_3 \, (Z\alpha)^2}
\nonumber\\[2ex]
&\sim& \frac{1}{2} \, (Z\alpha)^6 \,
\ln \frac{\epsilon_1}{(Z\alpha)^2} \,
\ln \frac{\epsilon_2}{(Z\alpha)^2} \,
\left(A_1 + A_3 - A_2 \right)\,,
\\[3ex]
\label{I3}
I_3 &=& \int_0^{\epsilon_1} {\rm d}\omega_1 \, \omega_1 \,
\int_0^{\epsilon_2} {\rm d}\omega_2 \, \omega_2 \,
\frac{1}{\omega_1 + \omega_2} \,
\frac{(Z\alpha)^2}{\omega_2 + A \, (Z\alpha)^2} 
\nonumber\\[2ex]
&\sim& - \frac{1}{2} \, (Z\alpha)^6 \,
\ln \frac{\epsilon_1}{(Z\alpha)^2} \,
\ln \frac{\epsilon_2}{(Z\alpha)^2} \, A^2\,.
\end{eqnarray}

\end{document}